# Group-Level and Personalized Optimization for the Insula and Hippocampus Focal Electric Field in Transcranial Temporal Interferential Stimulation: A Computational Study

Taiga Inoue[1], Naofumi Otsuru[2], and Akimasa Hirata[1,3]

1. Department of Electrical and Mechanical Engineering, Nagoya Institute of Technology, 466-8555, Japan
2. Department of Physical Therapy, Faculty of Rehabilitation, Niigata University of Health and Welfare, Niigata, Japan
3. Center of Biomedical Physics and Information Technology, Nagoya Institute of Technology, 466-8555, Japan

**Corresponding Author:**

Akimasa Hirata

Department of Electrical and Mechanical Engineering, Nagoya Institute of Technology

Gokiso-Cho, Showa-ku, Nagoya 466-8555, Japan

E-mail: ahirata@nitech.ac.jp

**Abstract:**

**Objectives:** This study evaluated the efficacy of transcranial temporal interference stimulation (tTIS) for focal stimulation of the insula and hippocampus, which are clinically relevant but anatomically challenging targets. Individualized and group-level electrode optimizations were compared to determine whether generalized montages can provide reliable targeting while reducing the modeling demands.

**Methods:** Sixty high-resolution anatomical head models (30 individuals and their mirrored counterparts) were constructed from T1- and T2-weighted magnetic resonance images. The electric field (EF) distributions were determined using the scalar-potential finite difference method. The electrode montages and current ratios were optimized to minimize the root-mean-square error between the simulated and target EF envelope (EFE) distributions. A stimulation threshold of 0.3 V/m was applied. Subsampling analysis was performed to estimate the number of head models required for stable group-level results.

**Results:** For insular targeting, a novel montage combining T7–P7 and Fp1–Fp2 achieved the highest focality. The focality was comparable to most individualized configurations and reduced interindividual variability. For hippocampal targeting, a newly proposed montage combining F7–T7 and T8–P8 yielded the best group-level focality. However, individualized optimization was required in a subset of cases to achieve adequate off-target suppression. Reliable group-level EF patterns were obtained using ~20 models for the insula and ~9 for the hippocampus.

**Conclusions:** The findings show optimal transcranial stimulation montages depend on the target's anatomical depth. For cortical targets, including deep areas like the insula, group-level montages derived from sufficiently diverse anatomical models can achieve both high focality and applicability. However, for subcortical targets like the hippocampus, individualized optimization is recommended to maximize focality and minimize off-target activation, despite requiring fewer models to achieve stable group-level patterns.

## 1. Introduction

Neuromodulation techniques, such as transcranial electric stimulation (tES) [1] and transcranial magnetic stimulation (TMS) [2], are widely used in clinical and research settings. These noninvasive approaches have been implemented in the treatment of neurological and psychiatric disorders [3] and are valuable tools for investigating the neural mechanisms underlying cognitive processes [4] and pain perception [5, 6].

In two established tES modalities, transcranial direct current stimulation (tDCS) and alternating current stimulation (tACS), a current is delivered via scalp electrodes to generate electric fields (EFs) in targeted brain regions. However, high cerebrospinal fluid (CSF) conductivity and interindividual anatomical variability can distort the EF distribution, often causing unintended activation near the target [7-10]. Computational modeling and related pipelines [11-13] offer advanced EF prediction and informed protocol optimization [14-16]. However, conventional tES remains largely limited to superficial cortical regions because EF intensity and focality decrease significantly with depth [17].

Transcranial temporal interference stimulation (tTIS) has emerged as a novel method for overcoming these limitations, offering the selective stimulation of deeper brain regions while minimizing EF exposure in more superficial tissues. This technique uses two high-frequency electrical currents (in the kHz range) with slightly different frequencies to generate a low-frequency envelope localized within the targeted areas [18-20]. Selective stimulation occurs via the low-pass filtering properties of neural membranes, a mechanism confirmed by both computational modeling and experimental validation [21]. tTIS can effectively modulate subcortical structures, including the hippocampus, thalamus, and pallidum, highlighting its superiority over tACS in reducing unintended stimulation in off-target regions [19, 22-25].

Rampersad, Roig-Solvas, Yarossi, Kulkarni, Santarnecchi, Dorval and Brooks [19] demonstrated improved focality compared with tACS but encountered electrode optimization difficulties. Studies in [22] and [23] both underscore the necessity of personalized montages for reliably targeting deep subcortical structures. Personalization has been demonstrated to enhance EF targeting and reduces variability for structures such as the thalamus, hippocampus, and caudate nucleus [22]. The importance of personalized optimization is emphasized for reliable hippocampal and pallidal stimulation using 25 head models [23]. Group-level and individually optimized montages were compared across 25 personalized models in [25]. Individualized optimization has been shown to substantially improve focality and EF strength for deep targets, with spatial shifts of up to 9.3 cm when using non-personalized configurations.

The insula, which is involved in pain processing, interoception, and emotional regulation [26-28], is a clinically important but challenging target due to its intermediate depth and surrounding CSF. This region exhibits abnormal activities in major depressive disorder [29], indicating the potential therapeutic value of targeted modulation. However, noninvasive stimulation studies on tDCS or tACS have encountered difficulties in effectively targeting the insula, even after optimization or with clarification of key factors [30, 31]. Furthermore, no computational studies have systematically examined its stimulation with tTIS. Given the anatomical characteristics of the insula, tTIS may offer more precise noninvasive modulation of this region, with our previous study demonstrating its pronounced efficacy at depths greater than 10–15 mm [24]. In addition, the hippocampus, which plays a central role in memory formation, learning, and spatial cognition, is a deeply embedded brain structure that is essential for the consolidation of short-term memory into long-term memory.

Here, we assessed the feasibility of tTIS for localized stimulation of the insula and hippocampus by directly comparing group-level and individualized electrode optimizations. We also determined the minimum number of anatomical head models required for robust group-level results, providing guidance on balancing targeting precision with computational resources. Furthermore, we quantified the performance reduction when applying an individual's optimized montage to other subjects, using focality metrics based on target-to-non-target ratios and EFE amplitude analysis to provide complementary insights into montage selection.

To achieve these aims, high-resolution computational modeling was conducted using 60 individualized head models derived from structural magnetic resonance imaging (MRI) data. Electrode configurations and injection current ratios were optimized using the scalar-potential finite difference (SPFD) method, and EFE distributions for both the insula and hippocampus were analyzed. We applied the same computational framework implemented in our previous optimization study for tDCS [32], with modifications for the tTIS-specific envelope field calculations.

## 2. Materials and Methods

The computational procedure used in the present study is almost the same as that described previously [32], in which the number of electrodes, their locations, and the injection current were optimized in tDCS.

### 2.1. Head Model Generation and Volume Conductor Model

Structural MRI data (T1- and T2-weighted, 1 mm voxel size, 3.0 T) were obtained from 22 male participants (21–55 years) registered in an open-access repository (NAMIC: Brain Multimodality; http://hdl.handle.net/1926/1687) and from 10 men (21–24 years) at Niigata University of Health and Welfare. Two datasets in NAMIC were excluded because of image artifacts or poor skull–CSF contrast.

In total, 60 computational models were generated, including 30 original and 30 mirrored heads. Each voxel-based model (0.5 mm resolution) was segmented into 16 tissues [7, 9], and conductivity values, representing the head as a volume conductor, were assigned as described previously [33]. Figures 1 (A) and (B) present the human head model and the classification of deep brain regions.

Each electrode was modeled as a combination of a conductive rubber segment and a sponge (Fig. 1 [C]). The rubber layer had a conductivity of 0.1 S/m and a thickness of 1 mm. It was embedded centrally within a 5-mm-thick sponge with a conductivity of 1.6 S/m, simulating saline-soaked conditions. The electrode was square-shaped with each side measuring 20 mm. Its surface was 60% conductive rubber.

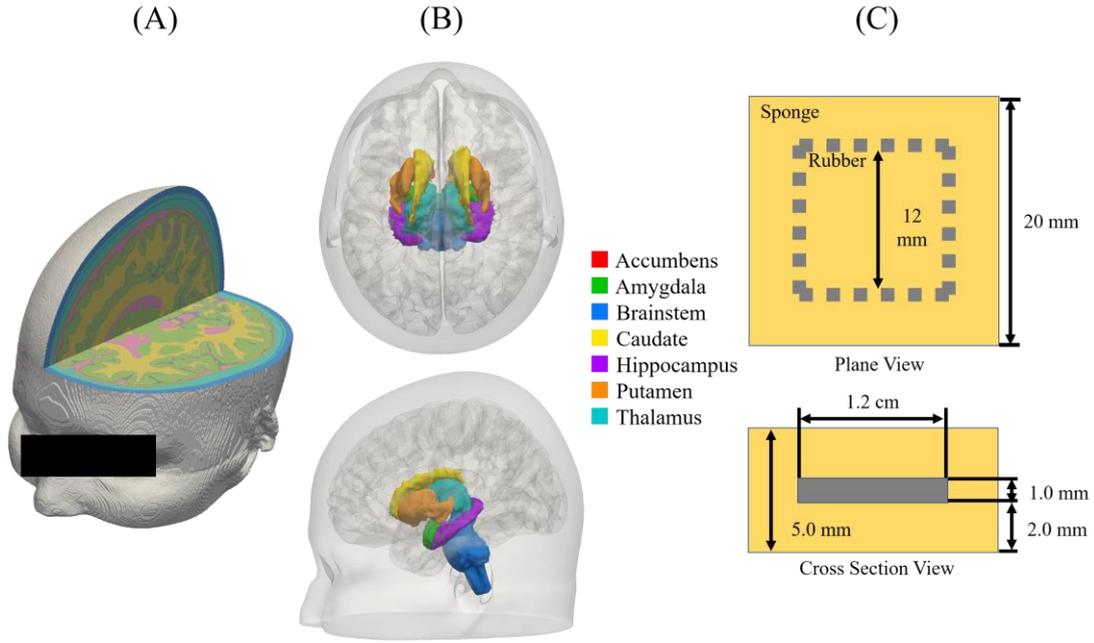

**Figure 1**. (A) Voxel-based cross-section of a human head model; (B) segmentation of deep regions in the head model; and (C) top and cross-sectional views of the electrode model.

**2.3. Scalar-Potential Finite Difference Method**

Under quasi-static approximation (<10 MHz), electromagnetic exposure can be divided into separate electric and magnetic field problems [34, 35]. The same principle is applied for direct current. The SPFD method [36] was employed to calculate scalar potentials at all nodes using the finite difference method, as follows:

$$\sum_{n=1}^{6} S_n \varphi_n - \left( \sum_{n=1}^{6} S_n \right) \varphi_0 = j\omega q \quad (1)$$

where $\varphi_n$ is the electric scalar potential at node $n$; $q$ is the stored charge corresponding to the injection current on the right-hand side term in the equation; $\omega$ is the angular frequency; and $S_n$ is the edge conductance obtained by voxel geometry and tissue conductivity [36].

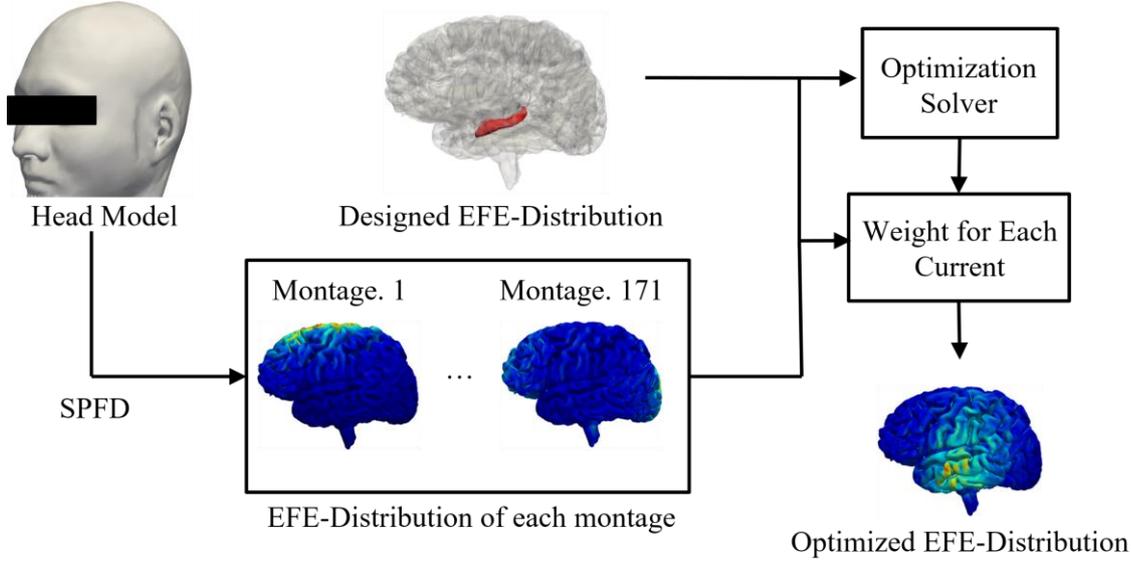

**Figure 2**. Flowchart for calculating the electromagnetic field for the optimized electric field envelope (EFE) computation in hippocampal targeting.

### 2.3. Optimization Procedure

The optimized injection current ratio for each montage was determined by aligning the computed EFE with the target EFE. For the insula, this alignment was focused solely on the cerebral cortex. For the hippocampus, alignment included both the cortex and deep brain regions (Fig. 2). The EFE distributions for all electrode pairs ($M = {}_{19}C_2$) from the 10 to 20 system (excluding the ground) were computed and stored on the $E_{Dataset}$, which represents the EF in the target region generated by external currents.

$$E_{Dataset} = [E_1 \cdots E_m] = \begin{bmatrix} E_1(r_1) & \cdots & E_m(r_1) \\ \vdots & \ddots & \vdots \\ E_1(r_N) & \cdots & E_m(r_N) \end{bmatrix} = \begin{bmatrix} E_{x1}(r_1) & \cdots & E_{xm}(r_1) \\ E_{y1}(r_1) & \cdots & E_{ym}(r_1) \\ E_{z1}(r_1) & \cdots & E_{zm}(r_1) \\ \vdots & \ddots & \vdots \\ E_{x1}(r_N) & \cdots & E_{xm}(r_N) \\ E_{y1}(r_N) & \cdots & E_{ym}(r_N) \\ E_{z1}(r_N) & \cdots & E_{zm}(r_N) \end{bmatrix} \quad (2),$$

where $N$ is the number of surface vertices in the target; $m$ is the number of electrode-pair combinations (m ≤ 171); and $E_m(r_1)$ denotes the EF vector at location $r_1$ induced by montage $m$. Only surface EF vectors were used to reduce the computational cost, which scaled quadratically with the model resolution.

The optimal current weighting coefficients were obtained by minimizing the root-mean-square error (RMSE) between the $E_{Dataset}$ and desired EFE using a constrained least-squares solver with limits on the current ratios to avoid potential excessive current injection. EF surfaces were registered to each individual's brain surface, as defined by FreeSurfer [37], and according to a previous method [38]. The simulations were performed on a workstation with 16 Intel® Xeon® CPUs (4 GHz), 256 GB RAM, and NVIDIA RTX A2000 GPUs.

### 2.4. Constrained Least Squares

The ideal focal EFE $E_{Dataset}$ can be expressed as a linear combination of the induced EFE distributions from

the two electrode pairs and their respective current ratios, as follows:

$$E_1 \times w_1 + E_2 \times w_2 = E_{Dataset} \times w = E_{Designed} \qquad (3),$$

where *w* denotes the current ratio for the *i*-th pair, which is derived using an optimization solver to minimize the RMSE in the desired EFE. To avoid potential side effects in clinical applications, each current ratio was constrained by the upper limit. Although this limit is set, higher injection currents in the kHz range are permissible because of the lower membrane response at these frequencies [39].

$$w_1 + w_2 = 2 \text{ mA} \qquad (4)$$

## 2.5. Target Region and Threshold for Stimulation

The insula and hippocampus are deep brain regions associated with depression and epilepsy and were thus selected as optimization targets [40, 41]. The input array $E_{Designed}$ for the optimization (Fig. 2) was defined as follows:

$$E_{Designed} = \begin{bmatrix} e_d(r_1) \\ \vdots \\ e_d(r_n) \\ \vdots \\ e_d(r_N) \end{bmatrix}, e_d(r_n) = \begin{cases} 1 & n \in T \\ 0 & n \in T^c \end{cases} \qquad (5)$$

Based on previous studies on tACS [6], the EFE threshold for neuromodulation was set at 0.3 V/m, the midpoint of the effective range commonly reported (0.2–0.5 V/m) and is corroborated by computational and in vitro evidence that link EFE values in this range to neuronal modulation. This threshold was selected to optimize the sensitivity for detecting neuromodulatory effects while minimizing the unintended stimulation of adjacent regions. EF distributions exceeding 0.3 V/m were identified as zones of effective stimulation.

## 2.6. Group-Level Analysis

Individual variability in brain morphology complicates the direct comparison of EFE distributions. To perform group-level evaluation, each individual model was spatially aligned to the MNI-ICBM 152 template [42] via affine transformation. The EFE distribution from each model was then projected onto the template by mapping each element to its nearest template element. Group-level maps were constructed by averaging the projected EFEs from all the 60 subjects.

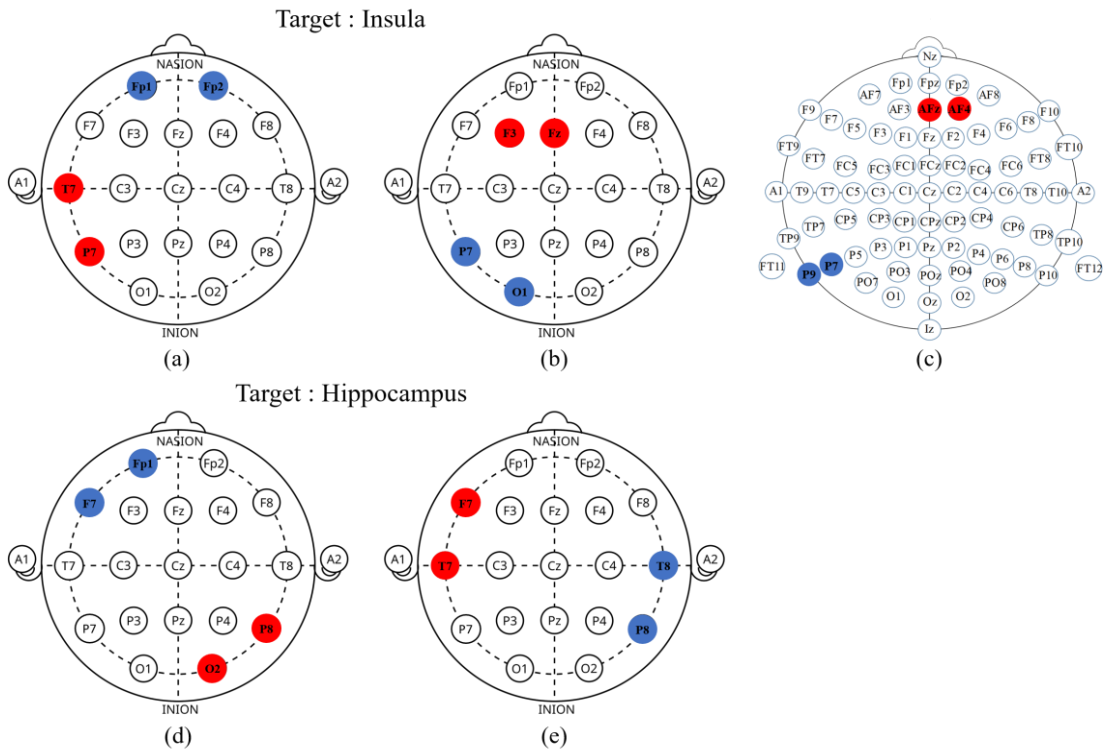

**Figure 3**. Electrode montages for insular and hippocampal stimulation. (a) Montage I (T7–P7 and Fp1–Fp2) and (b) Montage II (Fz–F3 and P7–O1) were frequently selected for insular stimulation in our optimization. For hippocampal stimulation, (c) Montage III (AFz–AF4 and P7–P9) used in a previous study, (d) Montage IV (F7–Fp1 and P8–O2), and (e) Montage V (F7-T7 and T8–P8) were frequently selected in our optimization. (f) The international 10-20 system.

To assess the effect of group size, the 60 models were randomly divided into two groups. For each group, the mean and standard deviation (SD) of the EFE values were calculated, and the correlation between the corresponding vertices was then determined. This process was repeated 500 times for each group size $N$ (2–30), and the resulting mean ± SD correlation coefficients were used to evaluate the influence of group size on EFE consistency.

## 3. Results
### 3.1. Optimization of Electrode Configurations for Insular and Hippocampal Stimulation

Across the 60 head models, the optimization process identified distinct montages for targeting the insula and hippocampus. As shown in Figure 3, Montage I (T7–P7 and Fp1–Fp2) was the most frequently selected (18/60) for the insula, followed by Montage II (F7–Fp2 and Fz–F3) (8/60). For the hippocampus, Montage IV (F7–Fp1 and P8–O2) was the most common (15/60), followed by Montage V (F7–T7 and T8–P8) (13/60). Given the anatomical proximity of the two targets and the limited number of insula-specific studies, a hippocampal montage reported in the literature was included for comparison. Each montage was optimized to minimize the evaluation function for its target. The required injection currents demonstrated interindividual variability in achieving the desired EFE distribution. The optimized injection current ratios were $I_1 = 0.911 \pm 0.196$ mA and $I_2 = 1.089 \pm 0.196$ mA for Montage I (insula), and $I_1 = 0.700 \pm 0.090$ mA and $I_2 = 1.300 \pm 0.090$ mA for Montage IV (hippocampus).

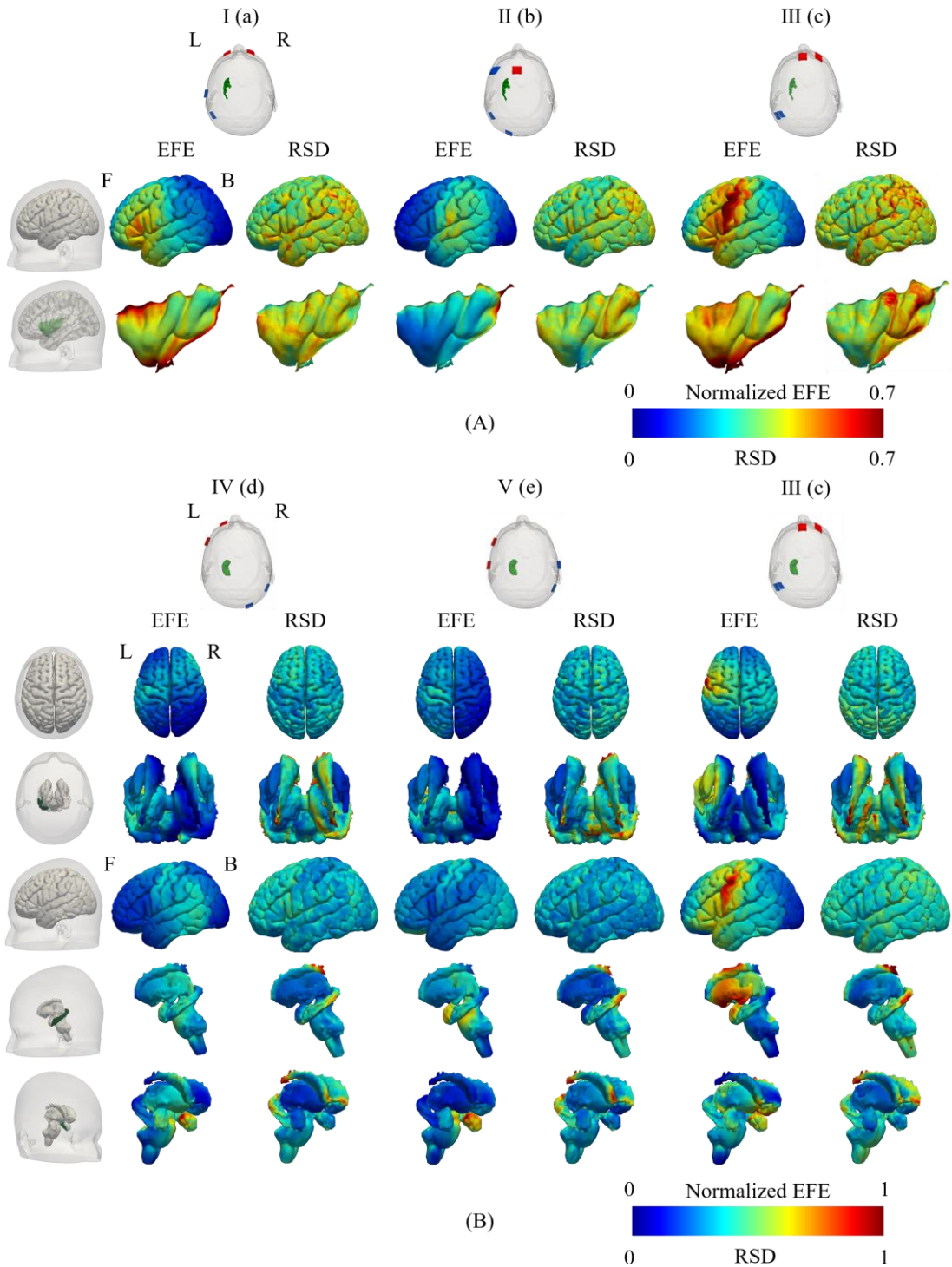

**Figure 4**. Mean ± relative standard deviation (RSD) for the electric field envelope (EFE) amplitude at the group level in the target and nontarget regions. Targeting (A) the insula and (B) hippocampus. The EFE amplitude was normalized to its maximum value within the target region.

### 3.2. Group-Level Validation

Based on the optimized electrode configurations identified in Section 3.1, an evaluation was conducted on the group-level effectiveness of targeting the insula using Montages I and II, which were the most frequently selected across models. The same evaluation was performed for the hippocampus using Montages IV and V. In each case, the injection current ratios were reoptimized for all individual models, and the mean values were adopted for group-level validation. The resulting current ratios were as follows: Montage I: $I_1 = 0.891$ mA and $I_2 = 1.109$ mA; Montage II: I $I_1 = 1.087$ mA and $I_2 = 0.913$ mA; Montage IV: $I_1 = 0.687$ mA and $I_2 = 1.313$ mA; and Montage V: $I_1 = 0.724$ mA and $I_2 = 1.277$ mA. For comparison, Montage III, which was taken from a previous study[23], was also evaluated, with the injection currents individually optimized for the head models in the present study. Consequently, the injection currents were $I_1 = 0.985$ mA and $I_2 = 1.015$ mA for the insula and $I_1 = 1.122$ mA and $I_2 = 0.878$ mA for the hippocampus.

Figure 4 shows the normalized EFE amplitudes projected onto a standard brain template. In the insula (Fig. 4A), Montages I and III produced high amplitudes across most of the region. However, Montage II elicited more localized stimulation, showing high amplitudes within a limited portion of the insula but not in other cortical areas. Montage III also induced comparably high amplitudes in the primary motor cortex, indicating reduced focality. In the hippocampus and adjacent accumbens (Fig. 4B), Montages I and II produced high amplitudes, whereas Montage III primarily activated the caudate and putamen and induced higher amplitudes in the cortical regions than in the deeper targets.

Figure 5 shows the normalized mean and maximum EFE amplitudes in the target and neighboring regions. In the insula (Fig. 5A), Montage I achieved high mean amplitudes confined to the insula, whereas Montage III produced comparable amplitudes in the middle temporal region, indicating activation of the non-target cortex. Montage II showed lower insular amplitudes and minimal differences with the surrounding areas. At maximum amplitude, Montages I and II maintained surrounding-region values at ~80% of the insula value, whereas Montage III exceeded this value in the superior temporal region. Figure 5(B) shows that the normalized mean EFE amplitude in the hippocampus is approximately the same between Montages IV and V. However, in

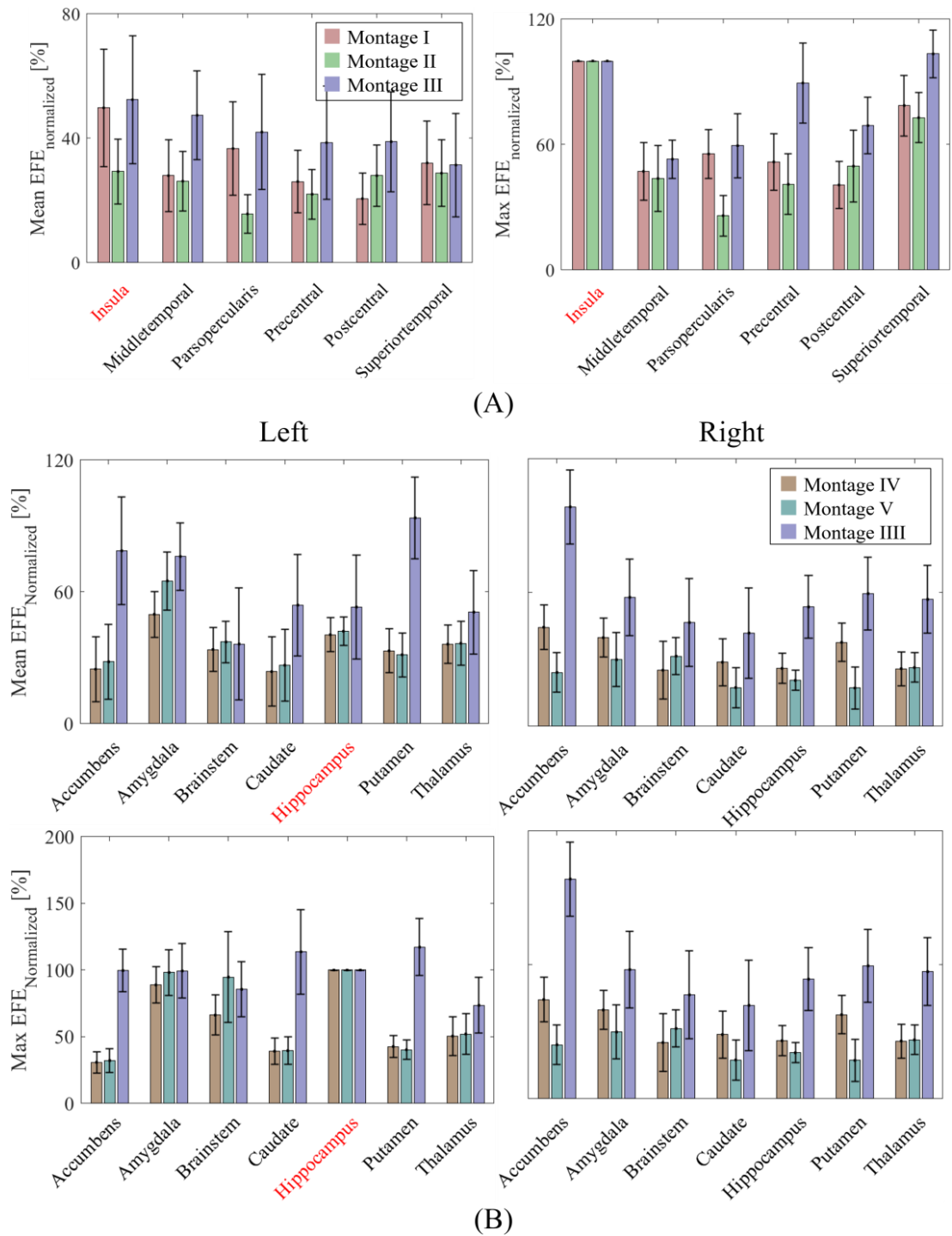

**Figure 5.** Mean and maximum electric field envelope (EFE) amplitudes at the group level in the target and non-target regions. Targeting the (A) insula and (B) hippocampus. The EFE amplitude was normalized to its maximum value within the target region.

Montage IV, higher mean EFE amplitudes were observed in the left amygdala and right accumbens compared with that in the hippocampus. Moreover, when comparing non-target regions between Montages IV and V, several regions showed clearly higher mean amplitudes in Montage IV (right accumbens, right amygdala, right caudate, and right putamen). Although both the Montages IV and V showed the highest maximum EFE amplitudes in the hippocampus, the trends in non-target regions were similar to those observed for the mean EFE amplitude.

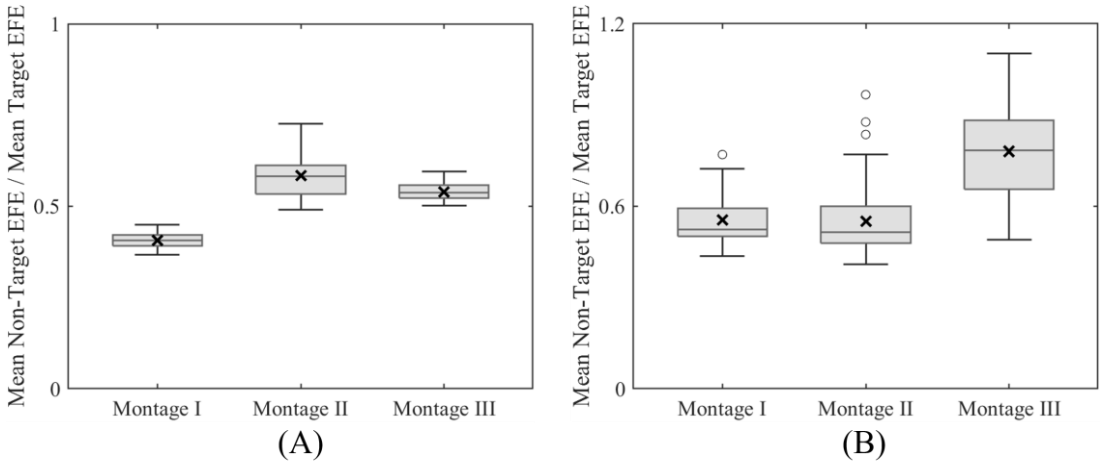

**Figure 6**. Evaluation of stimulation focality for each montage, measured as the ratio of the mean amplitude in non-target regions to that in the target region: Targeting the (A) insula and (B) hippocampus. (EFE: electric field envelope).

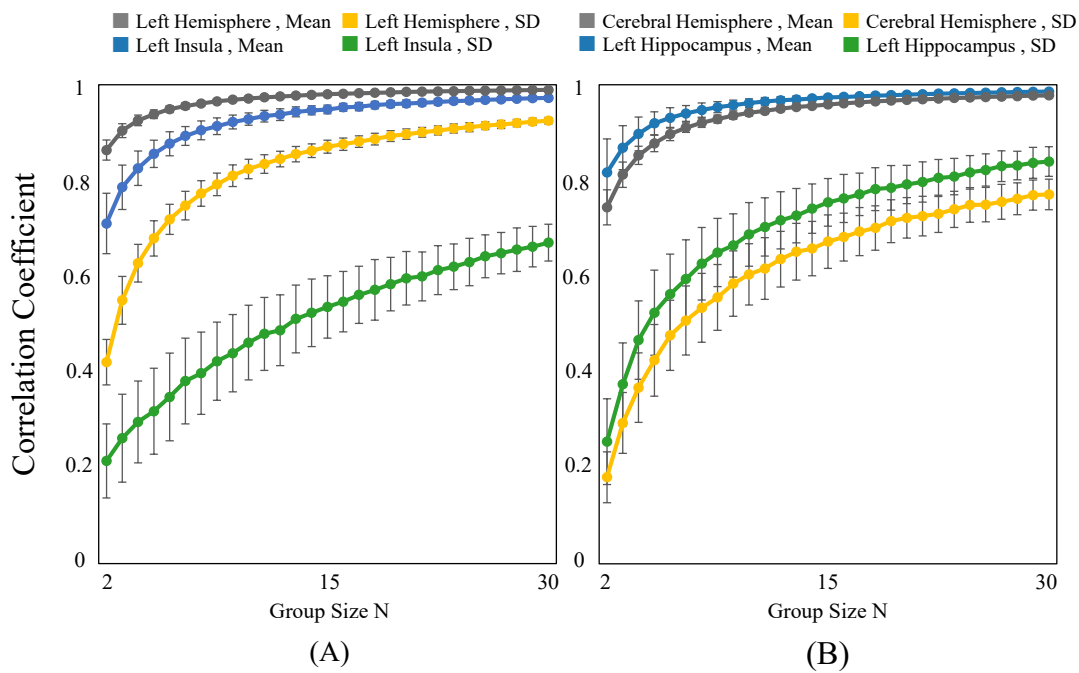

**Figure 7**. Effect of group size on the group-level EFE amplitude measured (mean ± SD) on the surface of the target region and non-target region for the following montages: (A) I and (B) V.

Mean EFE amplitude ratio of non-target to target regions indicate greater focality (Fig. 6). In the insula (Fig. 6A), Montage I showed a significantly higher focality than Montages II and III ($p < 0.05$), indicating that Montage I was the most effective among the tested configurations in restricting stimulation to the target region. In the hippocampus (Fig. 6B), the focality ratios for both the mean and maximum amplitude metrics were higher in the individually optimized configurations than those in Montage V.

Subsampling analysis was conducted for the Pearson correlation between random subgroups (N = 2–30; 500 iterations). Twenty or more models were sufficient for a reliable estimation (correlation ≥ 0.95) of the insular EF distribution (Supplementary Figure 7). For the hippocampus, approximately 9 models were sufficient for the target region and ~13 for the cortical non-target region. Calculating the SD required larger sample sizes owing to the interindividual variability in the EF distribution.

### 3.3. Comparison of Individual Optimization Results and Group-Level Effective Conditions

The effectiveness of the group-level electrode configurations (Montage I for the insular cortex and Montage V for the hippocampus) was evaluated against individually optimized configurations based on focality and EFE amplitude strength.

Focality, defined as the ratio of the mean and maximum EFE amplitudes in the target region to those in the nontarget regions, showed target-dependent differences (Supplementary Figure 8). For the insula, group-level Montage I achieved a higher focality based on the mean-ratio metric, whereas individually optimized configurations produced slightly higher focality based on the maximum ratio. These findings show that, although the group-level montage more consistently suppressed non-target stimulation, some individuals achieved better peak localization with individualized optimization. For the hippocampus, individually optimized configurations yielded a higher focality than Montage V for both metrics, indicating more consistent localized stimulation across subjects.

The strength of the EFE amplitude was also evaluated (Figure 9). A threshold of 0.3 V/m was applied, and the proportions of the target and surrounding regions that exceeded this threshold were determined. To ensure comparability across montages, the injected currents were scaled to a normalized value where 10%, as an example, of the target region exceeded the threshold. For the insula, scaling factors of 0.87 and 0.955 were applied for the individually optimized configuration and Montage I, respectively. For the hippocampus, scaling factors of 2.05 and 2.14 were applied for the individually optimized configuration and Montage V, respectively. These scaling factors better reflect the lower intrinsic EFE intensity for the hippocampus than for the insula and required larger adjustments to achieve comparable suprathreshold coverage in the target.

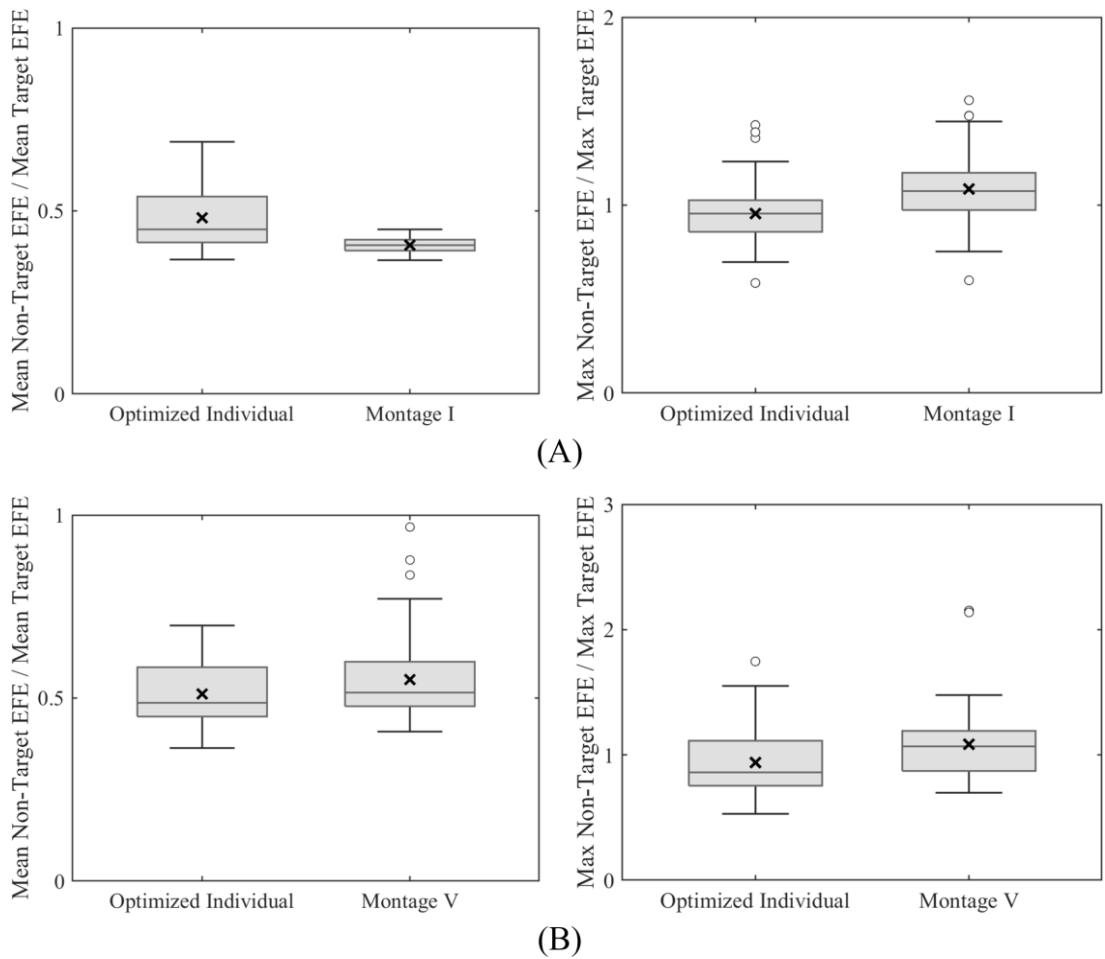

**Figure 8**. Evaluation of stimulation focality for each montage, assessed by calculating the ratio of both mean and maximum amplitudes in non-target regions relative to the target region: (A) the insula and (B) the hippocampus.

When targeting the insula (Figure 9A), the mean proportion of non-target regions that exceeded the threshold did not differ significantly between Montage I and the individually optimized configurations. However, larger variability was observed in the latter, particularly in outlier cases, indicating elevated suprathreshold proportions in some models across multiple regions. This variability indicates that although some individuals benefited from tailored montages, others experienced unintended increases in off-target stimulation.

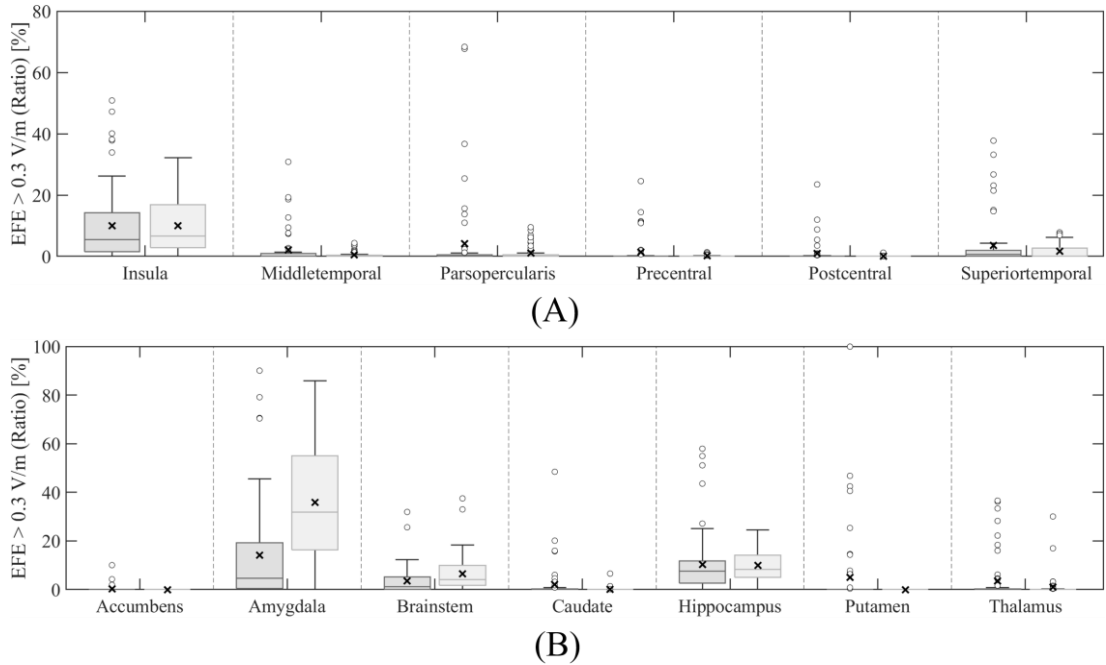

**Figure 9**. Ratio of electric field envelope amplitude exceeding the 0.3 V/m threshold in the (A) insula and (B) hippocampus. Injection currents were scaled before normalization to ensure that 10% of the target region exceeded the threshold.

When targeting the hippocampus (Figure 9B), the mean proportion of non-target regions above the threshold, especially in the amygdala, differed significantly between Montage V and the individually optimized configuration. Specifically, the individually optimized configuration generally produced lower off-target amplitudes in adjacent structures, further highlighting the benefits of personalization for deep-brain targets.

## 4. Discussion
### 4.1 Optimization of Electrode Configurations

This study is the first systematic evaluation of tTIS optimization for both the insula and the hippocampus, two clinically relevant but anatomically challenging targets. By comparing individualized and group-level optimization for these regions (Fig. 3), we identified target-dependent differences in the trade-off between focality and applicability across individuals without case-by-case optimization.

Compared with our previous work on superficial cortical targets, such as M1 and S1, using linear alignment montages [24], the insula, while also cortical, lies 10–20 mm deeper in the brain and is partially covered by surrounding lobes, creating access difficulties. This intermediate depth resulted in electric field distributions (Figs. 4A and 5A) that are distinct from those of superficial cortical targets. In contrast, the hippocampus is a subcortical target located even deeper, featuring distinct anatomical boundaries and a lower positional variability across individuals (Figs. 4B and 5B). These anatomical differences underpin the variation in the optimization outcomes observed in this study.

For the insula, the most frequently selected configuration (Montage I) achieved a high focality and lower interindividual variability (Fig. 6A), which is comparable to the results of individualized optimization in many cases, thereby demonstrating high applicability. For the hippocampus, although Montage V performed best at the group level, individualized optimization was necessary in some cases to achieve adequate off-target suppression (Fig. 6B), which reflects its lower applicability despite the anatomical consistency.

**4.2 Group-Level Validation and Model Requirements**

Subsampling analysis (Supplementary Fig. 7) revealed that approximately 20 anatomical models were required to obtain stable group-level maps for the insula, whereas ~9 models were sufficient for the hippocampus (~13 for the cortical non-target region) at a correlation coefficient threshold of 0.95. In comparison, our previous study on superficial cortical targets, such as M1 and S1 [24], required 13 models to achieve the same correlation level when the analysis was restricted to the target region.

The insula had higher model requirements than the M1/S1 to achieve robust group-level patterns, which reflects the complex interindividual variability in CSF thickness and sulcal morphology around this deep cortical target. In contrast, the hippocampus required fewer models, likely because of its more consistent position and shape in the Montreal Neurological Institute (MNI) space, which facilitates quicker group-level averaging. However, as discussed in Section 4.3, a lower model requirement for spatial stability does not necessarily imply that group-level montages can achieve optimal focality in the subcortical structures.

From a practical standpoint, these estimates guide the allocation of computational resources. Group-level designs for superficial cortical targets may require ~13 models, while deep cortical and subcortical targets require ~20 and ~9 models, respectively, to achieve stable target-region patterns. Nevertheless, the montage design should still reflect the anatomical and physiological characteristics of each target.

**4.3 Comparison of Individual Optimization and Group-Level Conditions**

When targeting the insula, the group-level montage (Montage I) achieved a higher focality in terms of the mean target-to-nontarget amplitude ratio (Fig. 6A), whereas individualized configurations yielded a slightly higher focality when assessed using the maximum ratio. This pattern indicates that group-level designs exerted more uniform suppression of off-target stimulation, whereas individualized designs occasionally achieved more concentrated peak amplitudes within limited insular subregions.

When targeting the hippocampus, individualized optimization outperformed Montage V in both the mean- and maximum-ratio focality metrics (Fig. 6B) and reduced off-target activation, particularly in adjacent structures, such as the amygdala (Fig. 9B). These findings are consistent with those in a previous work [25], which also emphasized the need for individualized optimization to improve focality and minimize off-target activation of subcortical targets. One possible explanation for the relatively higher amplitudes in the amygdala and nucleus accumbens is the limited number of mesh elements in these regions (1,394 and 776 elements, respectively) compared with the hippocampus (2,059 elements). This limitation, combined with the anatomical adjacency of the amygdala to the hippocampus, likely influenced the observed results. Furthermore, our identification of a novel group-level montage for the hippocampus offers a useful starting point and complements earlier studies by indicating that generalized designs may still offer practical value, despite higher preference for subject-specific optimization. Thus, although the hippocampus showed high spatial stability in group-level maps with fewer models, achieving optimal focality without subject-specific optimization remains challenging.

Clinically, these results imply that for cortical regions, including deep cortical areas such as the insula, group-level montages informed by a sufficiently diverse model set can achieve both high focality and applicability, making them suitable for early-phase clinical trials. In contrast, for subcortical targets, such as the hippocampus, individualized optimization is required to achieve the desired stimulation specificity and minimize the unintended activation of neighboring structures.

**4.4 Limitations and Future Directions**

This study relied entirely on computational modeling and did not perform in vivo validation of the stimulation effects. The conductivity values were obtained from literature-based data without accounting for subject-specific variations, which may have influenced the actual EF distribution. The subject pool also had a relatively narrow age and sex distribution, which potentially limits the generalizability of the findings. Moreover, only one optimization criterion (RMSE of a target EFE) was considered; alternative optimization metrics or multi-objective approaches may yield different montage designs. Future studies should integrate personalized conductivity estimation, physiological measurements, and functional outcome assessments to validate computational predictions and refine montage optimization strategies.

**5. Conclusions**

Group-level electrode montages can effectively stimulate certain cortical targets, such as the insula, with minimal unintended activation. However, individualized optimization appears advantageous for deeper and anatomically variable structures, particularly in the hippocampus, and may be essential for achieving the desired stimulation specificity. These findings underscore the importance of tailoring montage design to the anatomical characteristics of the target region and objectives of the stimulation protocol.

**Credit authorship contribution statement Taiga Inoue**: Writing – original draft, review & editing, Methodology, Investigation, Formal analysis, Data curation, Visualization. **Naofumi Otsuru**: Writing – review original draft, Investigation, Supervision, Project administration. **Akimasa Hirata**: Writing – review &editing original draft, Conceptualization, Formal Analysis, Methodology, Funding acquisition, Supervision, Project administration.

**Declaration of Competing Interest**

The authors have declared no competing or conflict of interest.

**Funding statement**

This work was supported by JSPS KAKENHI (grant numbers 21H04956 and 25K02972).